**Dynamics of a first order electronic phase transition in manganites**

T.Z. Ward[1], Z. Gai[1,2], H.W. Guo[1,4], L.F. Yin[3], J. Shen[3,4]

[1]Materials Science and Technology Division, Oak Ridge National Laboratory, Oak Ridge, Tennessee 37830, USA
[2]Center for Nanophase Materials Sciences Division, Oak Ridge National Laboratory, Oak Ridge, Tennessee 37830, USA
[3]Dept. of Physics, Fudan University, Shanghai 200433, China
[4]Dept. of Physics and Astronomy, The University of Tennessee, Knoxville, Tennessee 37996, USA

* To whom correspondence should be addressed: shenj5494@fudan.edu.cn, wardtz@ornl.gov

By reducing an electronically phase separated manganite $(La_{1-y}Pr_y)_xCa_{1-x}MnO_3$ single crystal thin film to dimensions on the order of the inherent phase domains, it is possible to isolate and monitor the behavior of single domains at a first order transition. At this critical point, it is possible to study the coexistence, formation and annihilation processes of discrete electronic phase domains. With this technique, we make several new observations on the mechanisms leading to the metal insulator transition in manganites. We observe that domain formation is emergent and random, the transition process from the metallic phase to the insulating phase takes longer than the reverse process, electric field effects are more influential in driving a phase transition than current induced electron heating, and single domain transition dynamics can be tuned through careful application of temperature and electric field.

## I. Introduction

Electronic phase separation (EPS), or electronic inhomogeneity, in complex oxides has been linked to colossal magnetoresistance, the Mott transition, multiferroicity and high $T_c$ superconductivity; however there is very little experimental information related to how exactly the electronic domains seed, grow and transition. While often associated with soft matter systems, such as polymers, the inherent complexity of strongly correlated materials is owed to the energetic overlaps of spin-charge-lattice-orbital parameters. The most widely held theories attempting to explain EPS include complex electronic interactions driven by random, self-organized strain distributions or local chemical disordering that act as energetically favorable seeding points[1-9]. The mechanisms that drive EPS and the rich phase diagrams present in many complex oxide materials are considered to be the result of strong electronic correlations[1,10-16]; however understanding how exactly these parameters interact to form phase coexistence has been elusive. The ability to observe the phase formation process and recognize how external stimuli drive discrete phase dynamics is a promising means of furthering our understanding of these strongly correlated systems. To address this issue, we will discuss work using a technique that allows us to probe a single or few electronic domains as they undergo phase transition with a high time resolution.

EPS in complex materials can have phase domains ranging from nanometers to micrometers[17-19]. By reducing the size of the material to the length scale of the inherent phase domains, it is possible to reduce the transport channel in a manner that forces the probing electrons to interact with regions of high resistance due to domain blockage across the channel similar to a serial resistor network. Changes to the inherent resistivity of the confined region arising from electronic phase formation and transition will have a dominant signature in transport measurements[20-24]. (LaPrCa)MnO$_3$ is of particular interest for these studies as it can exhibit extremely large coexisting charge ordered insulating (COI) and ferromagnetic metal (FMM) phase domains of over 1μm[10]. This makes reducing these systems using traditional wet etch photolithography a viable means to reach these length scales.

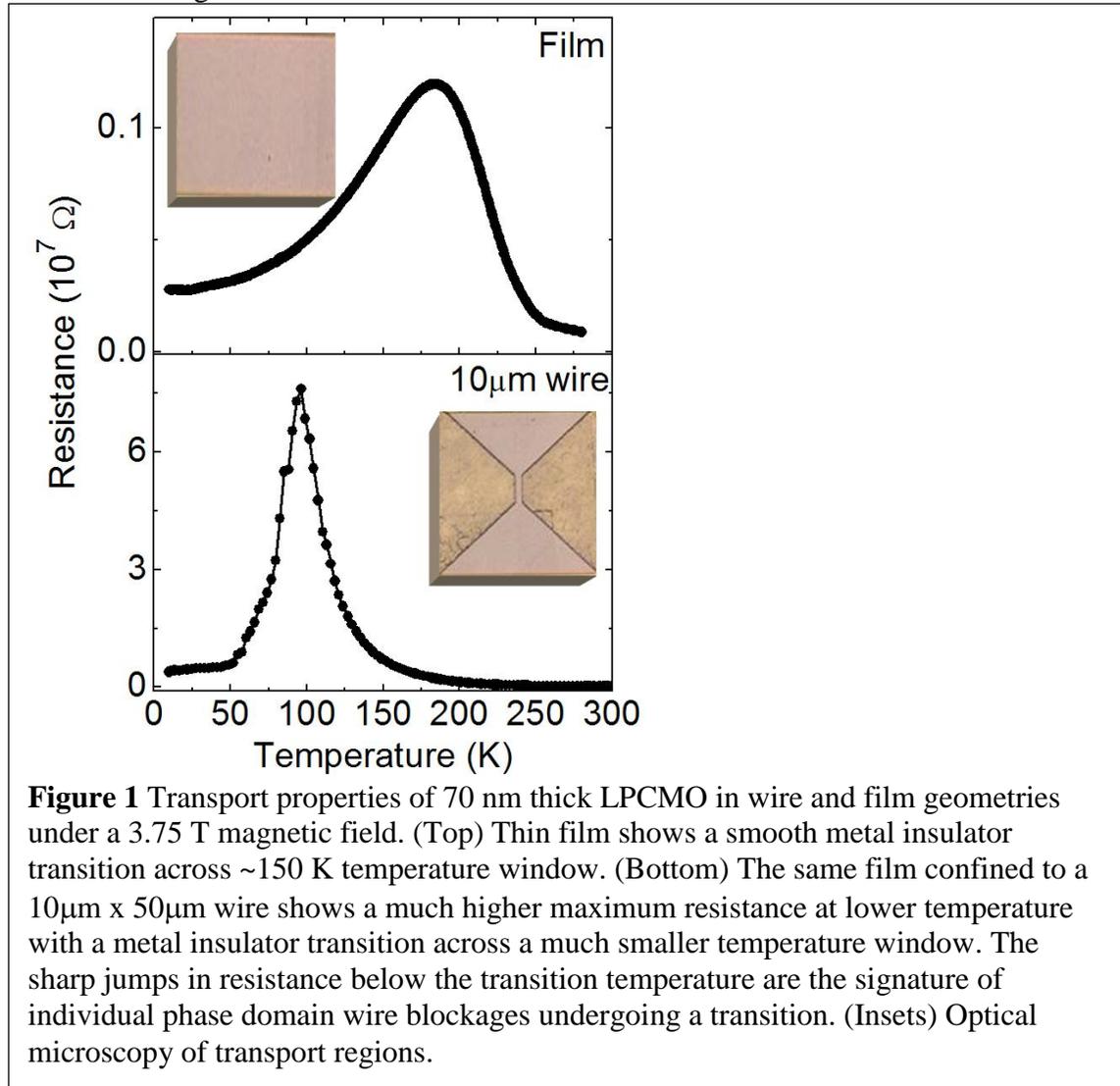

**Figure 1** Transport properties of 70 nm thick LPCMO in wire and film geometries under a 3.75 T magnetic field. (Top) Thin film shows a smooth metal insulator transition across ~150 K temperature window. (Bottom) The same film confined to a 10μm x 50μm wire shows a much higher maximum resistance at lower temperature with a metal insulator transition across a much smaller temperature window. The sharp jumps in resistance below the transition temperature are the signature of individual phase domain wire blockages undergoing a transition. (Insets) Optical microscopy of transport regions.

Figure 1 shows a comparison of the temperature dependent resistance of an LPCMO thin film before and after confinement with a constant 3.75 T applied magnetic field to ensure ferromagnetic alignment. Before confinement, the film exhibits a typical behavior with a smooth metal-insulator transition (MIT) across a 150 K window. After reducing the film to a 10μm wide wire, the resistive behavior changes dramatically. In this geometry, the transport channel is blocked by the formation of insulating domains within the wire which dominate the resistive signal. This pushes the device's maximum resistance orders higher and pushes the MIT temperature lower. Immediately below the transition temperature, there is a dramatic, stepped drop in the resistance as dominant individual domains in the transport channel transition to the metallic phase. By fixing temperature and magnetic field at the critical point of the stepped transition in the wire geometry, it is possible to identify and characterize individual phase domains. The discrete jumps in resistance observed in the wire geometry between 75 K and 95 K are not visible in the film geometry as the probing current follows the path of least resistance along the large transport network thus rendering single domain transitions invisible (fig. 1). In this work, we will apply this technique to gain a deeper insight into the formation and transition processes at play in these systems.

## II. Experiments

Thin films of $(La_{1/2}Pr_{1/2})_{5/8}Ca_{3/8}MnO_3$ (LPCMO) with a thickness of 70 nm were grown on $TiO_2$ surface terminated $SrTiO_3$ substrates with a miscut angle 0.1° using laser MBE (248 nm, 1 J/cm$^2$ fluence). Growth was done in an ultrahigh vacuum system (UHV) with a base pressure of $< 1 \times 10^{-10}$ Torr. RHEED was used to insure layer-by-layer growth. The substrates were held at a constant 820° C in a flowing $O_2$ environment at 1 mTorr. After growth, samples were post-annealed at 800° C in 1 atm flowing $O_2$ for 10 hours and then slowly cooled to room temperature to fill any oxygen vacancies. X-ray diffraction showed single phase epitaxial coverage and AFM measurements showed clean, terraced surfaces[25]. Magnetization measurements were performed using a Quantum Design Magnetic Properties Measurement System (MPMS). Transport measurements were conducted in a Quantum Design Physical Properties Measurement System (PPMS) with a base temperature of 1 K and magnetic field control of +/- 9 T. Due to the high resistances of the samples and the necessary measurement frequencies, a Keithley 2400 power supply (8μs full current recover time) was used as a current source with a National Instruments DAQ USB-6281 (625 kHz measurement rate) to measure voltage. Constant current measurements were all conducted at 500 nA unless otherwise stated. These were controlled by homebuilt LabView drivers for data collection. The relatively high resistances of the materials allowed for 2-probe measurements though 4-probe was also tested and showed no appreciable difference in results. Etching was accomplished using standard wet-etch photolithography in a 10% KI solution. Post-etch microscopy showed total film removal around desired device geometries.

## III. Results and Discussion
### A. Thermal cycling effects on electronic domains

By holding the 10 μm x 50 μm x 70 nm LPCMO wires at a critical point in temperature and magnetic field, it is possible to isolate a single electronic domain as it undergoes a transition[20-22]. Cycling the temperature far above this critical point forces the material to return to its paramagnetic insulating (PI) parent phase. If electronic domain seeding occurs at defect sites or long-range quenched disorder in the lattice structure, it would be expected that the same resistive behavior would be seen on each return to the same temperature when following identical thermal cycling procedures, because these large energy pinning sites would be present on each cycle. We observe that this is not the case. Figure 2 shows resistance as a function of time for four thermal cycles where each cycle is described as 300 K to 10 K to 83 K (20 minutes of transport data taken) to 300 K at a rate of 5 K/minute; the current source was left on at all times. Here, 10 second (450,000 measurements taken at 45 kHz) increments are averaged for each point. Across the 20 minute scanning times, each cycle shows very different resistive levels from the other cycles, is stable within its own range, and does not collapse to a common cycle independent state. Cycles 1 and 3 have ~90% difference in resistance while cycles 1 and 4 have ~10% difference. In the widely accepted percolative transport model, this means that the transport channel is changing with each cycle and, therefore, upon each thermal cycle the seeding of electronic domains is different. We can also observe that there does not appear to be evidence of a memory effect where the FMM phase seeds more easily after subsequent thermal cycles, since the resistive level is not clearly associated with cycle order, i.e. cycle 4 has a lower resistance than cycle 1 but higher resistance than cycles 2 and 3. These findings point to an emergent seeding mechanism independent of large energetic pinning due to surface/edge roughness or long-range quenched disorder.

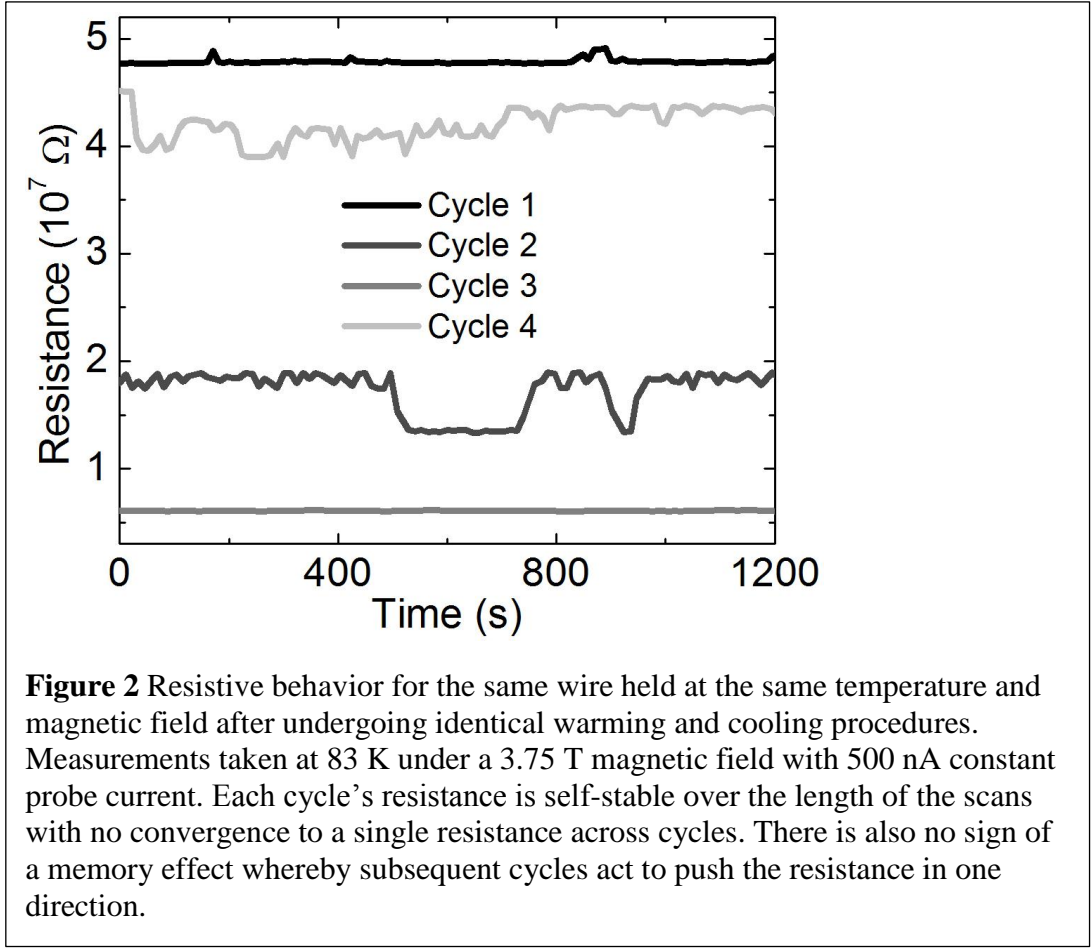

**Figure 2** Resistive behavior for the same wire held at the same temperature and magnetic field after undergoing identical warming and cooling procedures. Measurements taken at 83 K under a 3.75 T magnetic field with 500 nA constant probe current. Each cycle's resistance is self-stable over the length of the scans with no convergence to a single resistance across cycles. There is also no sign of a memory effect whereby subsequent cycles act to push the resistance in one direction.

We also observe that the stability and dynamic behavior of each cycle is different. To more closely investigate this we compare 10 seconds of high resolution resistance vs time data taken randomly from each of the 4 cycles at 45 kHz [figure 3]. In cycles 1, 2 and 4 we see clear indications of single domain transitions as regular resistive jumps occur[20,26]. The regularity of the resistive levels within each of the cycles are consistent with a percolative transport network in which a single electronic phase domain transitions between charge ordered insulating (COI) and ferromagnetic metallic (FMM) phases thereby creating discrete resistive levels. As described in previous work, this is consistent with a complex switching network in which a single electronic phase domain can be actively observed in different background states[20]. With each cycle we observe new dynamics within the wire where discrete fluctuation levels, phase transition probabilities and background resistive states are different. This offers strong evidence that the formation of electronic phase domains are seeded through an emergent process and not dominated by long-range quenched disorder or defects which would act as pinning sites thereby create similar transport channels with similar active domains on each thermal cycle.

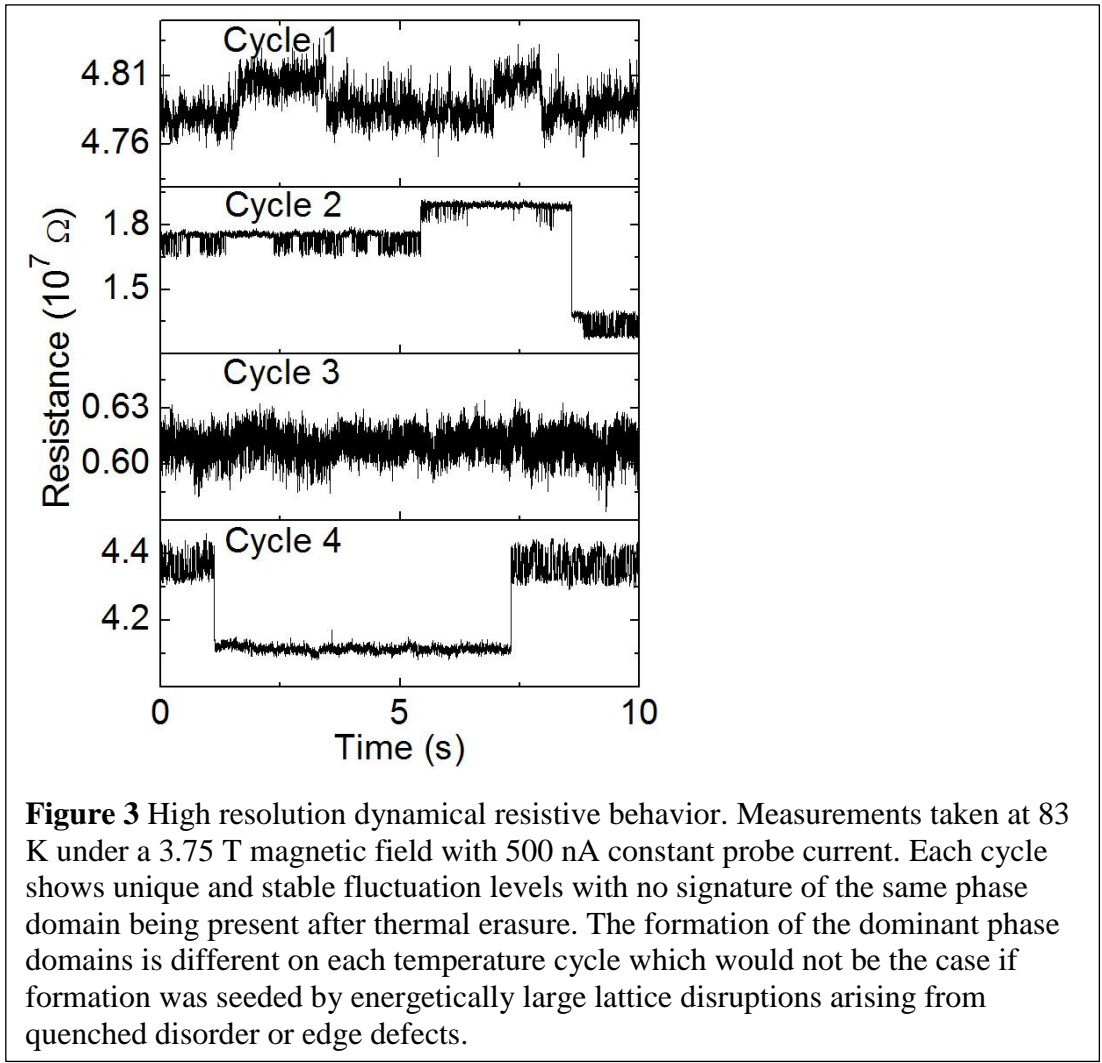

**Figure 3** High resolution dynamical resistive behavior. Measurements taken at 83 K under a 3.75 T magnetic field with 500 nA constant probe current. Each cycle shows unique and stable fluctuation levels with no signature of the same phase domain being present after thermal erasure. The formation of the dominant phase domains is different on each temperature cycle which would not be the case if formation was seeded by energetically large lattice disruptions arising from quenched disorder or edge defects.

To better understand the differences in behavior for each cycle, we investigated the binned data of all recorded high resolution measurements taken for each cycle across a 20 minute period (figure 4). Cycle 1 has a single extremely robust small domain active throughout. Cycle 2 shows a clear three level fluctuation with an imbedded two state fluctuation active on each level. Cycle 3 is the only cycle that shows no evidence of isolated single domain phase transitions—the spread in signal can be attributed to overlapping background fluctuations outside of the wire geometry. Cycle 4 possesses the most complex dynamical behavior, showing multiple stable states and imbedded two state fluctuations. From this, we can surmise that the stability of domains is not dependent on thermal cycling order. This again strengthens the argument that electronic domain formation and transition is an emergent phenomenon and not related to phase seeding at defect sites where local energetics would act to pin formation of phases to single locations in the transport chain with similar size and behavior on each thermal cycle.

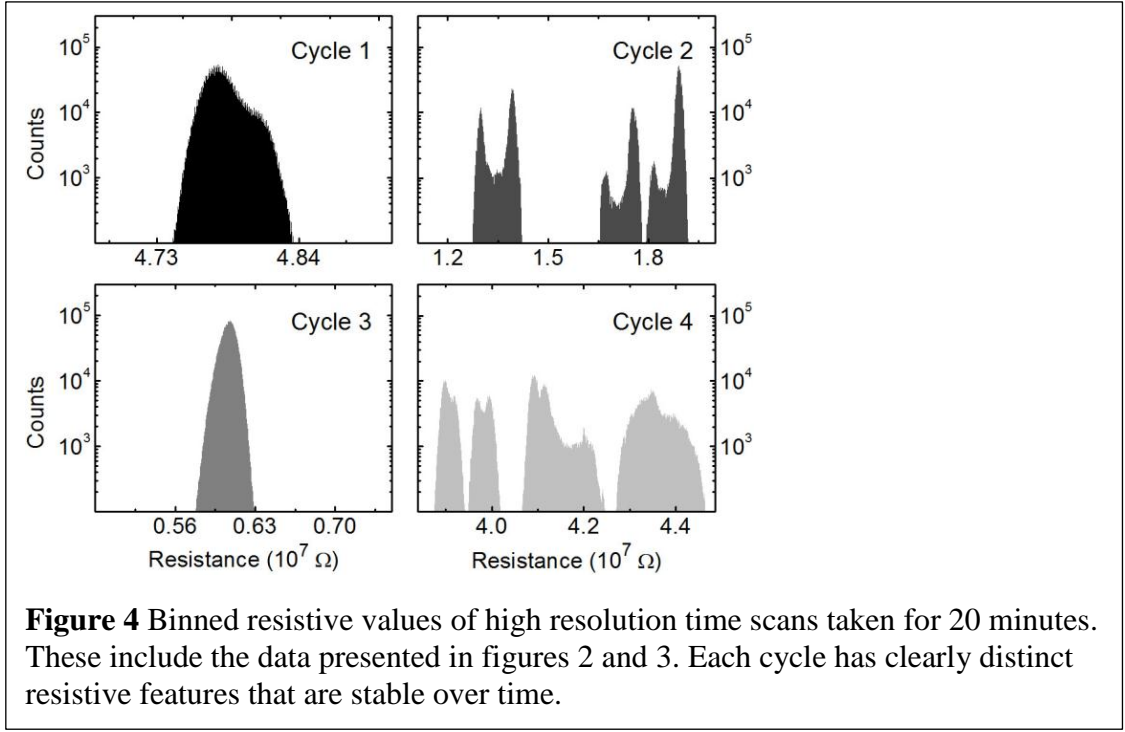

**Figure 4** Binned resistive values of high resolution time scans taken for 20 minutes. These include the data presented in figures 2 and 3. Each cycle has clearly distinct resistive features that are stable over time.

## B. Transition times of a single electronic phase domain

As shown above, confining a thin film to a wire of similar scale to the phase domains residing within the material creates an environment in which discrete regions can dominate a transport signal. This allows us to gather high resolution dynamical information on a single or few electronic domains residing in the wire as they transition between insulating and metallic phases. The behavior and response of a first order phase transition has not been well studied in previous works, but as we seek to better understand the mechanisms that drive electronic phase transitions and begin to implement these materials into practical devices, such as for magnetic cooling or memory applications, it is vital to investigate the timescales at which these transitions occur.

The data collected in cycle 2 from the above section is further investigated in this section, as it contains a single two state fluctuation that is present in three background states. This allows us to study the influence of background effects on a single electronic domain's dynamics. Figure 5 shows the resistance of a 10 μm wide wire held at a critical point of 83 K under a 3.75 T magnetic field. We see that there are three main resistive levels that are driven by the transition between electronic phases. Of particular importance, is that within each of these three background levels there is a single active domain that can be observed. This means that any differences in this single domain's dynamics can be directly attributed to changes in the applied electric field arising from background resistive changes, local current effects as the resistive network is modified, and correlations between neighboring domains as they undergo transition. We interpret the high and low resistance states as arising from a single phase domain set in a complex switching network dominating the transport signal as it transitions between the ferromagnetic phase to the charge ordered phase[20,22,27]. The inset shows a typical transition from the FMM phase to the COI phase followed a few milliseconds later by a transition back to the FMM phase. Note that this same transition behavior is seen on each of the three levels and is not an artifact of the power supply which has an orders faster response time and would show a linear response in an RC circuit due to constant current mode setting.

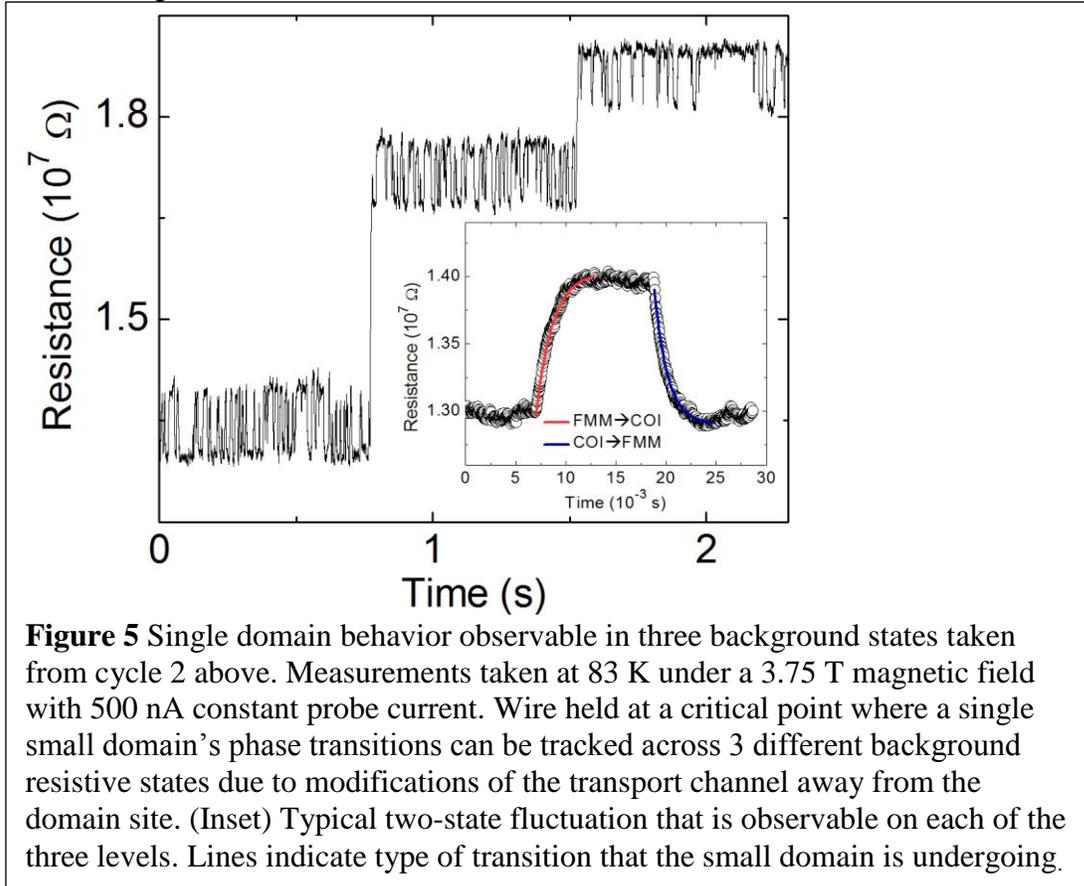

**Figure 5** Single domain behavior observable in three background states taken from cycle 2 above. Measurements taken at 83 K under a 3.75 T magnetic field with 500 nA constant probe current. Wire held at a critical point where a single small domain's phase transitions can be tracked across 3 different background resistive states due to modifications of the transport channel away from the domain site. (Inset) Typical two-state fluctuation that is observable on each of the three levels. Lines indicate type of transition that the small domain is undergoing.

To find the transition lifetime of the single electronic domain as it undergoes transition, we use an inverse exponential fitting: $R = A*e^{-t/\tau}$ where R is resistance, A is a constant, t is time and $\tau$ is transition lifetime. Thousands of transitions on the 3 levels were fitted; the results of which are shown in figure 6 with the error bars representing standard deviation in fitted lifetimes. The larger standard deviations observed in the lower $\Delta R$ transitions are due to fewer observed transitions which will be discussed in more detail below. The first thing that we can observe is that the metallic to insulating transition time is consistently longer than the insulating to metallic transition. This behavior was repeatable on subsequent thermal cycling with the disordering transition always averaging the shorter transition time. This could be a response to the disordered FMM phase requiring a longer time to order itself into the COI phase than for the COI phase to disorder to FMM. The observed transition times are on the order of milliseconds which is several orders longer than what would be expected for a purely quantum electrical transition or from thermal dissipation resulting from free energy release[28,29]. This longer transition time may be the signature of an intermediate glassy phase and are quite different from the nanoseconds transitions that have been induced using vibrational excitations[30] or the femtoseconds to picoseconds transitions induced through photo excitations[31]. These short transition times have been described as forced transitions where the probing excitation directly separates and melts the charge/orbital order parameters from the larger system[32]. This is an important point as it suggests that phase transitions that are driven through external stimulation push the systems far out of equilibrium so should not be considered when trying to understand the dynamics of a phase transition at a stable critical point.

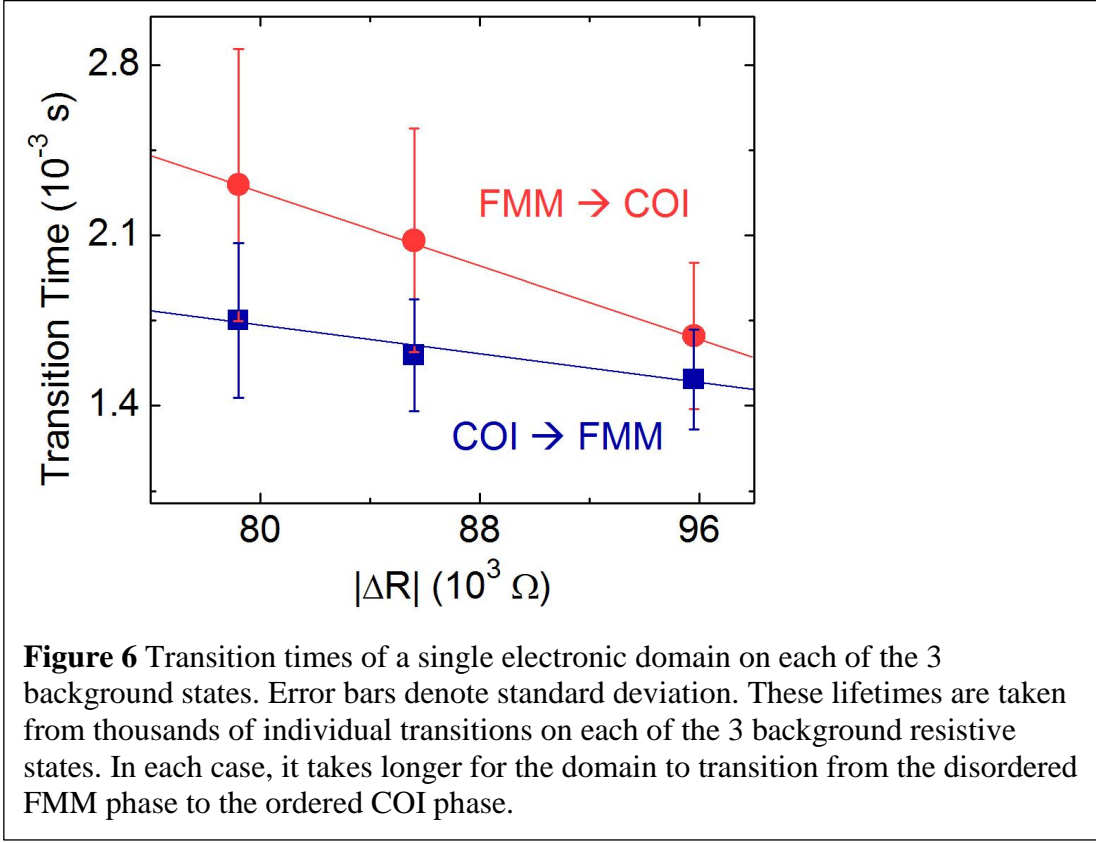

**Figure 6** Transition times of a single electronic domain on each of the 3 background states. Error bars denote standard deviation. These lifetimes are taken from thousands of individual transitions on each of the 3 background resistive states. In each case, it takes longer for the domain to transition from the disordered FMM phase to the ordered COI phase.

There is however a difference in the attempt frequency for domain transition and a subtle shift in the apparent magnitude of the resistance for the constantly active domain as a function of the network's resistive state. There is a difference in the number of transitions on each level; for all data analyzed, the higher resistance background showed 1118 transitions, the mid-level showed 2191 transitions and the low level showed 6529 transitions with equal time sections on each level. This means that the attempt frequency of the low level is much higher than at the higher resistive background states. We also note that when the background network is in its lowest resistance state, the small fluctuator shows a change in resistance ($\Delta R$) of 96 k$\Omega$ as it oscillates between metallic and insulating. When the background network is in its highest resistance state, the small fluctuator shows a $\Delta R$ of 79 k$\Omega$ between phases. Each background state represents a slight difference in the transport network, with the lowest resistive background state having the lowest applied electric field and the lowest local current density due to having more open transport channels[20]. Previous works investigating how current and E-field influence percolative transport relied on looking at global effects through cycling the current and field which are unable to give a local view of exactly how single domains are influenced[33-36]. However, using the confinement technique it is possible to investigate exactly how local current effects and electric field effects influence single domain behavior. The differences in the contribution to the resistive signal and the shift in transition attempt frequency means that the higher applied electric field and higher current density applied to the small domain when in the higher background resistive state influences domain stability. It is not clear whether this change in ΔR is due to a shift in the geometry of the network or is a more fundamental change in the volume of the domain itself where local heating could either disorder a small region in the COI phase or act to order the FMM phase. We can surmise however that the higher field and current density act to stabilize the small domain, as its transition rate is far lower in this regime. To better understand how the percolative network and local domains are influenced, we must look at the effects of small changes in temperature to domain dynamics and the effects of varying current.

**C. Heating effects on formation and stability of single phase domains**

When trying to understand the formation and stability of phase domains, it is important to recognize how changes in the percolation network can drive single domain dynamics. As discussed above in the case of a single domain active in three background states, the mechanism at play in changing the dynamics of a single domain could be attributed to changes in the applied electric field arising from changes to the larger network, local current effects resulting from changes to the current path, or to inter-domain correlation effects. The present experimental setup makes direct imaging and investigation of cross-domain correlations impossible to study; however we can address the question of thermal effects arising from changes in the current density across the network in the form of local Joule heating at the current path. To do this, we observe how changes to the applied current influence the formation of domains and compare that to how small changes in sample temperature influence the properties of a single active phase domain.

Spurred by resistive random access memory applications (RRAM), the manganites have been well studied in regards to E-field induced resistive switching though there is still a great deal of debate as to the precise mechanism[33,36-39]. In these studies, the insulating phase melts to the metallic phase under applied voltages with the proposed mechanisms ranging from thermal effects arising from Joule heating, ion migration at the sample interface, or modification of the transport channel through dielectrophoresis. Interestingly, the vast majority of the work in this area use extremely high electric fields and therefore high applied currents. It is for this reason that there is some debate as to whether E-field or current are the most important in driving a resistive switch. By isolating a single electronic phase domain and investigating its response to current and temperature, it is possible to add some new insight into this controversy.

In figure 7, a 10 μm wide LPCMO wire was cycled from 300 K to 10 K to 80 K under a 3.75 T magnetic field. After a 20 minute settling time, constant current resistance measurements were taken using 10 nA, 50 nA, 100 nA, 200 nA, 400 nA, and 500 nA with no thermal cycling between measurements. At each current setting, 20 minutes of data were taken at 45 kSamples/s followed by a 5 nA/s increase in current to the next level. While these applied currents and their driving voltages are well below previously observed cutoff limits said to drive a resistive switch in LPCMO,[37,40] the first observation that we can make from figure 7 is that even modest changes to the applied current can have a dramatic effect on the wire's resistance. We see >85% decrease in resistance when increasing current from 10 nA to 500 nA. We also see no clear breakpoint at which an applied current will completely transition a percolative network through the material leading to a global switching effect. Instead, we observe that there is an overall nonlinear response that is driven by small single-phase domains independently transitioning (figure 7 inset). The high resolution time scans show that by adjusting the current levels it is not only possible to activate individual domains within the material but to then drive their transition attempt frequency. As current is increased, a single domain's preferred state will go from insulating to metallic. This can be seen most clearly when observing the activity at 200 nA and 400 nA. We can compare fluctuations between these two driving currents by looking for a common resistive signature of the same phase domain transitioning in time. To do this, we compare the percent change in the resistance between the high resistive ($R_{high}$) and low resistive states ($R_{low}$) for each driving current. Considering that the active domain is a serial blockage of identical size regardless of driving current where applied current effects will influence all parts of the sample equally, the percent resistive change between high and low resistive states should be the same if the same region is transitioning. The 200 nA resistance plot has a very clear 2 level fluctuation with a ($R_{high}$-$R_{low}$)/$R_{low}$ of 13% that can be attributed to a single active domain transitioning between metallic and insulating with ~70% probability of being in the metallic state. The 400 nA resistance plot also shows a single active domain with a 13% ($R_{high}$-$R_{low}$)/$R_{low}$ signature but with ~99% probability of being in the metallic state. With increased current and higher driving electric field, we see a preference toward the metallic state. To answer the question of whether this melting of the insulating phase is due to thermal effects from Joule heating or is a direct result of the change in the electric field we must observe how changes in temperature influence the electronic phase preference.

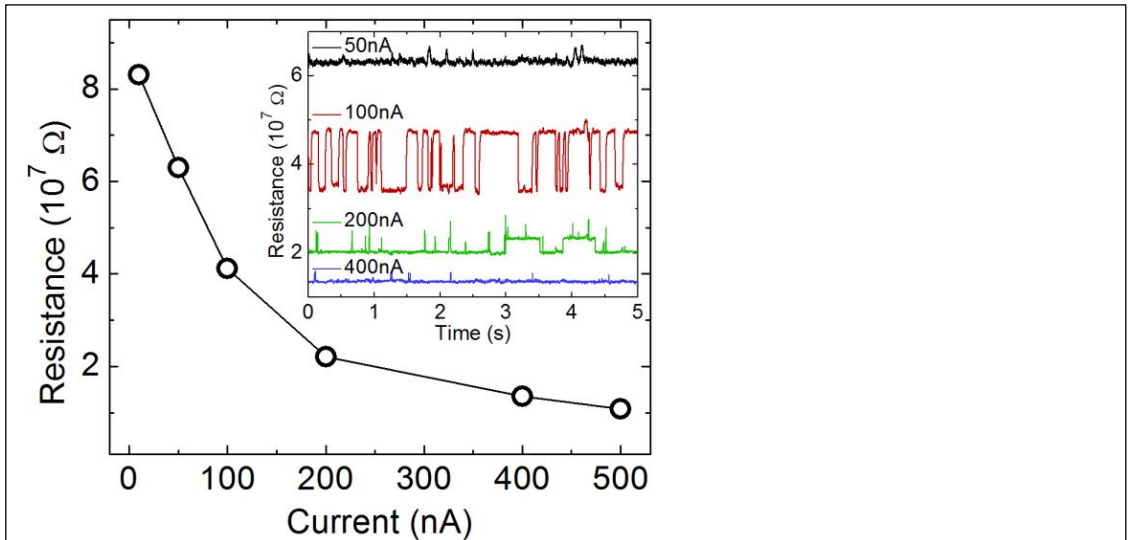

**Figure 7** Current effects on formation of electronic domains. Average resistances across time for LPCMO wire under increasing constant current while wire is held at 80 K under a 3.75 T magnetic field. There is no visible breakpoint at which the device switches at some critical current. (Inset) Typical high resolution resistance scans show that single domains can be activated and/or have their preferred phase tuned by changing the applied current.

To understand whether Joule heating is the driving factor in melting the insulating phase, we place the LPCMO wire near a critical point for a single domain transition following the thermal cycling procedure discussed above. Figure 8 compares the low time resolution resistance plots at three progressively higher temperatures with the corresponding binned values of the high resolution data given for clarity. At each of the presented temperatures we see a clearly active 2 state fluctuation that can be attributed to the same phase domain transitioning between metallic and insulating. At 80.3 K, there is a metallic preference of ~94%. If we are to assume that Joule heating was the cause of the reduction in resistance discussed above, an increase in temperature should necessarily increase the preference of the domain to a metallic phase. Instead, we see that by increasing the temperature in 0.1 K increments the domain's preference of phase shifts toward the insulating phase. At 80.4 K, the metallic phase is only present 73% of the time. While at 80.5K, the domain shows an almost exclusive preference for the insulating phase, spending only 3% of the time in the metallic phase.

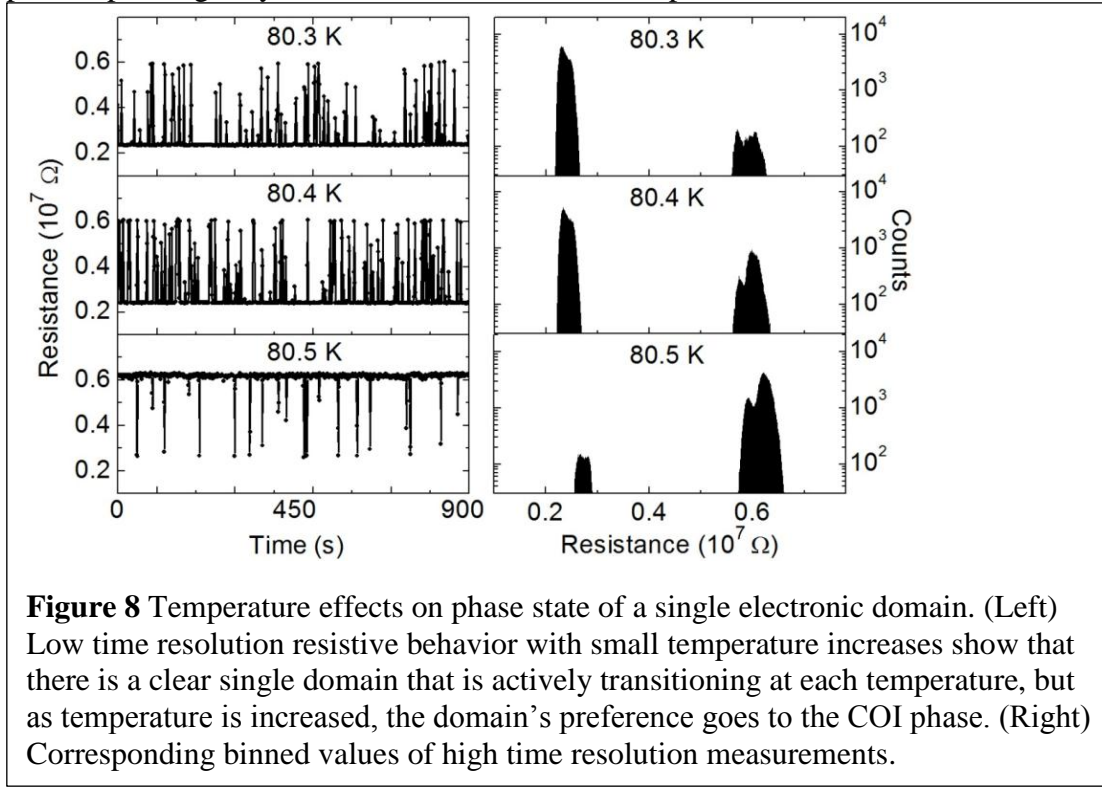

**Figure 8** Temperature effects on phase state of a single electronic domain. (Left) Low time resolution resistive behavior with small temperature increases show that there is a clear single domain that is actively transitioning at each temperature, but as temperature is increased, the domain's preference goes to the COI phase. (Right) Corresponding binned values of high time resolution measurements.

These findings suggest that in the percolation model, heating effects are not a significant factor in causing the transition from insulator to metal for low driving currents. This is surmised from the fact that while an increase in driving current and its necessary applied electric field increase the Joule heating in the sample, we see a preference for the metallic phase as the applied field is increased. When the probing current is held constant and increases in sample temperature are introduced environmentally, we see that a discrete domain prefers the insulating phase. Thus, it appears much more likely that it is indeed electric field effects which are the major driving force in pushing domains from an insulating to metallic phase as purely thermal effects are shown to push phase domains to the insulating phase.

## IV. Conclusion

By reducing the size of an electronically phase separated manganite to the scale of the phase domains present in the material, it is possible to observe one or a few phase domains at their critical point in a metal-insulator transition. This is possible due to the fact that in this regime the reduced transport network size forces the current path across otherwise resistively hidden regions. We observe that the formation and dynamics of electronic phase domains are not dependent on local physical disruptions and are in fact seeded in a non-repeatable emergent process. The transition time for a domain to disorder from the COI phase to the FMM phase is shown to be shorter than for the same domain to order from the FMM phase to the COI phase. Measurements taken as a single domain fluctuates between the FMM and COI phases show that it is possible to control the preference of a single domain's state by tuning applied electric field or system temperature. The influence of electron heating in forcing an insulator to metal transition is also shown to be negligible; the apparent driving mechanism for switching behavior can be attributed to the applied electric field. These findings should also find practical use, as many new functional applications, such as RRAM and magnetic cooling, rely on complex materials' inherent electronic inhomogeneity.

**Acknowledgments**
This effort was supported by the US DOE Office of Basic Energy Sciences, Materials Sciences and Engineering Division, through the Oak Ridge National Laboratory (TZW and HWG). A portion of this research was conducted at the Center for Nanophase Materials Sciences, which is sponsored at Oak Ridge National Laboratory by the Scientific User Facilities Division, US DOE. We also acknowledge partial funding supports from the US DOE Office of Basic Energy Sciences, the US DOE grant DE-SC0002136 (ZG), and the National Basic Research Program of China (973 Program) under the grant No. 2011CB921801 (LFY and JS).